\documentclass[10pt]{article}
\usepackage{latexsym}% for Box in \proof
\newtheorem{theorem}{Theorem}
\newtheorem{lemma}[theorem]{Lemma}
\newtheorem{corollary}[theorem]{Corollary}
\newcommand\QED{\ifhmode\allowbreak\else\nobreak\fi\quad\nobreak$\Box$\medbreak}
\newenvironment{proof}{\par\noindent{\bf
Proof:}}{\rm\enspace{\QED\par}}
\newcommand\commentout[1]{}
\begin{document}
\begin{titlepage}
\title{The Expected Size of the Rule k Dominating Set}
\author{
Jennie C. Hansen\\
\small Actuarial Math and Statistics Department\\
\small
Herriot-Watt University\\
\small J.Hansen@ma.hw.ac.uk\\
\\
 \and
Eric Schmutz\\
\small Department of Mathematics\\
\small Drexel University\\
\small Philadelphia, Pa. 19104 \\
\small Eric.Jonathan.Schmutz@drexel.edu\\
 }

\date{\today}
\maketitle
\abstract{
 Dai, Li, and Wu proposed Rule k, a  localized
 approximation algorithm that attempts to find a small connected
dominating set in a graph. Here we consider the \lq\lq average
case\rq\rq performance of Rule $k$ for the model of random unit
disk graphs constructed from $n$ random points in an $\ell_n\times
\ell_n$ square.
  If $k\geq 3$ and $\ell_{n}=o(\sqrt{n}),$ then the expected size
  of the
  Rule $k$ dominating set is $\Theta(\ell_{n}^{2})$ as
  $n\rightarrow\infty.$
If $\ell_{n}\leq \sqrt{{n\over 10\log n}},$
 then expected size of the
  {\em minimum} CDS is also $\Theta(\ell_{n}^{2}).$
 }

\vskip.5cm \noindent
{\bf keywords and phrases}: {\em  dominating
set, localized algorithm, approximation algorithm, performance
analysis, probabilistic analysis, Rule k, unit disk graph, }

\end{titlepage}
\section{Introduction}

In this paper we consider the problem of finding a small connected
dominating set for a {\em unit disk graph} $G=(V,E)$, where the
vertex set, $V$, is a set of points in $\Re^{2}$. Given the vertex
set $V$, the edge set $E$ is determined as follows: an {\it
undirected} edge $e\in E$ connects vertices $u,v\in V$  (and in
this case we say that $u$ and $v$ are adjacent) iff the Euclidean
distance between them is less than or equal to one. Unit disk
graphs have been used by many authors
 as simplified mathematical
models for the interconnections between hosts in a wireless
network, and random unit disk graphs have been used as stochastic
models for these networks. e.g. \cite{CCJ},\cite{Gilbert},
\cite{GuptaKumar},\cite{Hale},\cite{M},\cite{M2}.  We particularly
mention the work of the Hipercom Project, e.g.
\cite{J},\cite{JLMV}, because it is closely related to our work.
\par
A {\sl dominating set} in any graph $G=(V,E)$ is a subset ${\cal
C}\subseteq V$  such that every vertex $v\in V$ either is in the
set ${\cal C}$, or is adjacent to a vertex in ${\cal C}.$ We say
${\cal C}$ is a {\em connected dominating set } if ${\cal C}$ is a
dominating set and the subgraph induced by ${\cal C}$ is
connected. Obviously $G$ cannot have a connected dominating set if
$G$ itself is not connected. We use the acronym \lq\lq CDS\rq\rq
for  a dominating set ${\cal C}$ such that the subgraph induced by
${\cal C}$ has the same number of components as $G$ has. In this
paper we consider a {\em random} unit disk graph model, ${\cal
G}_{n}$, which is connected with asymptotic probability one. So,
in this case, any CDS for ${\cal G}_{n}$ will also be connected
with high probability.

 The identification of a small connected dominating set for
the graph which represents the  network is an important step in
several routing methods. The general idea of CDS-based algorithms
is to select a small  CDS, and have only those nodes responsible
for determining routes \cite{BD},\cite{BDS},\cite{SSZ},
\cite{WuLi}. It is believed that, by reducing the number of such
nodes, CDS-based algorithms
 reduce interference between transmitters in the same region
 and alleviate a related set of problems known collectively
as \lq\lq broadcast storm\rq\rq \cite{NTCS}. Furthermore the cost
of finding and maintaining routing information is smaller because
fewer nodes are involved. However it is beyond the scope of this
paper to consider
 {direct} measures of the utility of a small CDS  after it has
been found. In this paper we consider only a single  measure of
the algorithms' effectiveness, namely the size of the CDS it
finds.

Even with this simple measure of performance, there are
non-trivial algorithmic and analytical problems. It is an NP-hard
computational problem to find the {\em minimal} connected
dominating set in a unit disk graph \cite{Licht}. Hence there is
considerable practical interest in designing good approximation
algorithms for finding small connected dominating sets. See, for
example
\cite{AWF},\cite{CCCD},\cite{Span},\cite{GK},\cite{PL},\cite{SSZ}.
There have been various efforts to evaluate CDS algorithm's
average case performance using simulations. However, with the
exception of the theoretical parts of
\cite{BJ},\cite{J},\cite{JLMV},
 we are not aware of any
probabilistic analysis  that is proved mathematically.

 In this paper  we analyze `Rule $k$' ($k\geq 3$), a
 family of localized approximation algorithms proposed by
 Dai, Li, and Wu \cite{DaiWu},\cite{WuLi}. For each $k$, Rule $k$
 attempts to find a small CDS.
  We first choose an appropriate probability
model. Then, in the context of the model, we prove explicit
asymptotic bounds on the expected size of the dominating set that
Rule k produces. Thus our contribution is not the algorithm
itself, but rather a mathematically sound analysis of the
algorithm.

 Before describing Rule $k$, we
introduce some notation. We assume that each vertex has a unique
identifier taken from a totally ordered set. For convenience, when
$|V|=n$, we will use the numbers $1,2,\dots ,n$ as IDs, and will
number the vertices accordingly.
 If $x_i$ is any vertex, with ID given by $i$, let ...
 let $N(x_i)$ be
 the set consisting of $x_i$ and any vertices that are
 adjacent to $x_{i}.$
The CDS constructed by the Rule $k$ algorithm is denoted ${\cal
C}_{k}(V)$, and its cardinality is $C_{k}(V)=|{\cal C}_{k}(V)|.$
The elements  of ${\cal C}_{k}(V)$ are called \lq\lq gateway
nodes\rq\rq. ${\cal C}_{k}(V)$ consists of all vertices $x_{i}\in
V $ that are not excluded under the following version of Rule k:

 \vskip.5cm \noindent {\bf Rule k: } {\sl Vertex
$x_{i}$ is excluded from ${\cal C}_{k}(V)$  iff
 $N(x_i)$ contains at least one set of $k$ vertices
$x_{i_1},x_{i_2},\dots x_{i_k}$ such that
\begin{itemize}
\item    $i_{1}>i_{2}> \cdots
>i_{k}>i, $ and
\item The subgraph induced by $\bigl\lbrace x_{i_{1}}, x_{i_{2}},
\cdots ,x_{i_{k}}\bigr\rbrace $ is connected, and \item
$N(x_i)\subseteq \bigcup\limits_{t=1}^{k}N(x_{i_{t}}).$
\end{itemize}}
 \vskip.5cm\noindent
 Wu Li and Dai proved  that ${\cal C}_{k}(V)$ is a CDS, and
 they conjectured that the Rule $k$ dominating set is, in some
 sense,
 small on average.
 The main result in  this paper is a proof of their conjecture.
\par
 The rest of this paper is organized as follows. In the next
section we specify the  model and define the random unit disk
graph, ${\cal G}_n$. In Sections 3 we prove a local coverage
theorem that is needed in section 4 to prove an upper bound for
$E(C_{k}(V))$ Finally, in the remainder of the paper, we discuss
lower bounds and optimality issues. The appendix deals with a
related algorithm called the Marking Process.

\section{ Choice of Models}
Before estimating the expected size of the
 Rule $k$ dominating set, we must  specify
the underlying probability model. For any  real number $\ell >1$,
let ${\cal Q}(\ell)$ be an $\ell\times\ell$ square in $\Re^{2}.$
The particular choice of a square will be immaterial, but its size
will be very important. Let $\Omega_{n,\ell}={\cal Q}(\ell)
\times{\cal Q}(\ell)\times...\times{\cal Q}(\ell)$ be the $n$-fold
product space with the usual product topology.
 For each $n\geq 1$, let
${X}_{n,\ell,1}, {X}_{n,\ell, 2},\dots ,{X}_{n,\ell, n}$
  be a sequence of random points selected independently
  from a uniform
distribution on ${\cal Q}(\ell)$ and let ${\bf P}_{n,\ell}$ denote
the uniform probability measure on $\Omega_{n,\ell}$ induced by
the random variables ${X}_{n,\ell,1}, {X}_{n,\ell, 2},\dots
,{X}_{n,\ell, n}$.
 Finally, let ${\cal G}(n,\ell)$ be the
random unit disk graph with vertex set $V_{n,\ell}=\{
X_{n,\ell,1}, X_{n,\ell,2}, ...., X_{n,\ell,n}\}$ that is formed
from these vertices by putting an edge between two vertices iff
the Euclidean distance between the two vertices is less than or
equal to one.

We want to estimate the \lq\lq average\rq\rq  size of ${\cal
C}_{k}(V_{n,\ell})$ for large networks. As it stands, the expected
value $E_{n,\ell}(C_{k})$ $[=E(C_k(V_{n,\ell}))]$ is defined with
respect to the probability measure ${\bf P}_{n,\ell}$ on
$\Omega_{n,\ell}$ and depends on both $n$ and $\ell.$ We shall not
however attempt any multivariate asymptotic estimates. Instead, we
choose a suitable sequence, $\langle
\ell_{n}\rangle_{n=1}^{\infty}$, and consider the expected value
$E_{n,\ell_{n}}(C_{k})$ with respect to ${\bf P}_{n,\ell_{n}}$ as
$n\to\infty$. To simplify notation throughout, we will (usually)
suppress the dependence on the choice of a sequence $\langle
\ell_n\rangle_{n=1}^{\infty}$. Thus we write ${\cal G}_{n}$
instead of
  ${\cal G}(n,\ell_n)$, and write
 $E_n(C_{k})$ instead of $E_{n,\ell_{n}}(C_{k}).$
 Suppressing even $n$, we write ${\cal Q}$ instead of
  ${\cal Q}(\ell_{n}),$ and ${\bf P}$ instead of ${\bf
  P}_{n,\ell_{n}}.$

Conditions on the growth rate of $\ell_{n}$ will be clear from the
statements of theorems. However, to provide some perspective on
our choice of growth rates for $\ell_n$, we mention that it is
known that  the threshold for connectivity is $\ell_{n}=
\Theta(\sqrt{{n/ \log n}})$; if $\ell_{n}$ grows faster than this,
then the random unit disk graph ${\cal G}_{n}$
 will be disconnected with
probability $1-o(1)$ as $n\rightarrow\infty$. In this case, with
high probability, ${\cal C}_k(V_{n,\ell})$ will not be a {\it
connected} dominating set for ${\cal G}_{n}$. More precise
versions of these remarks are provided in the new book by
Penrose\cite{Penrose} which gives an up to date survey of  random
geometric graphs.
\par

Finally, throughout the remainder of this paper we adopt the
following notation. For  any points $p$ and $q$   in $\Re^{2},$
let $d(p,q)$ denote the ordinary Euclidean distance between $p$
and $q$ in $\Re^2$.

\section{Local Coverage by $k$ vertices}
 The next lemma is a purely geometric result which we require for
the proof of Theorem 2. To state the lemma, we need some
 notation. Let $\delta={1\over 2}-{\sqrt{3}\over 4}=.0669\dots,$ and define
$\rho$ by $\rho+2\delta=1;$ $\rho={\sqrt{3}\over 2}=.866\dots.$
Let\ $p$ be any point in $ {\cal Q}$, and let
$D^{'}=D_{1}(p)\bigcap {\cal Q}$ be the set of points in the
square ${\cal Q}$ whose distance from $p$ is one or less.
\begin{lemma}
\label{ksectors}
 There exist points
$z_{0},z_1,z_2\in D^{'}$ such that the following two conditions
are satisfied: \begin{itemize}\item for $s=0$ and $s=1$,\
 $d(z_{s},z_{s+1})\leq 1-2\delta$
 \item $D^{'}\subseteq
\bigcup\limits_{s=0}^{2}D_{\rho}(z_{s}).$
\end{itemize}
\end{lemma}
\begin{proof}
Consider first the case where $D_{1}(p)\subseteq {\cal Q}$, i.e.
$p$ is a point that is not near the boundary of the square. We
may, without loss of generality, choose the coordinate system such
that $p=(0,0)$ and such that the axes are parallel to the sides of
the square ${\cal Q}$. For $s=0,1,2,$ let $S_{s}$ be the sector of
$D_{1}(p)$
 consisting of those points
whose polar coordinates $(r,\theta)$ satisfy $r\leq 1$ and
${(2s-1)\pi\over 3}\leq \theta \leq {(2s+1)\pi \over 3}.$ Let
$z_{s}$ be the point in $S_{s}$ whose polar coordinates are
$({1\over 2},{2\pi s\over 3}).$ Then the first condition is
satisfied:  $d(z_{s},z_{s+1})=\sin{\pi\over 3}= 1-2\delta.$ It is
also straightforward to check that for $s=0,1,2$, $ S_{s}\subseteq
D_{\rho}(z_{s})$ and so the second condition is satisfied.

 Now consider the remaining case where $D_{1}(p)$ meets the
boundary of ${\cal Q}$.  Choose points $z_{0},z_{1},z_{2}$ as
before so that $D_{1}(p)\subseteq
\bigcup\limits_{s=0}^{2}D_{\rho}(z_{s})$ and $d(z_{s},z_{s+1})\leq
1-2\delta.$  We are not done because one or more of the points
$z_{s}$ may not lie in ${\cal Q}.$ In particular, if $z_s\notin
{\cal Q}$, then there is a (unique) $z_s'\in {\cal Q}$ such that
$d(z_s, z_s')=\inf\{d(z_s,z): z\in{\cal Q}\}$. We replace $z_s$ by
$z_s'$ and observe that every point of $D^{'}$ is closer to
$z^{'}_{s}$ than it is to the original point $z_{s}$. Hence
$S_{s}\bigcap {\cal Q}\subseteq D_{\rho}(z^{'}_{s}).$ After
replacing all$z_s$ such that $z_s\notin{\cal Q}$ by the
corresponding $z_s'$  we obtain three points that satisfy the
conditions of the lemma.
\end{proof}
\noindent Fix $k\geq 3,$ the $k$ in \lq\lq Rule k\rq\rq. Suppose
$m$ points $P_1,P_2,\dots ,P_{m}$ are selected independently and
uniform randomly in $D^{'}(p)$.
 Let
${\cal K}_{m}$ be the event that, for some $1\leq
i_0<i_1<i_{2}<\dots <i_{k-1}\leq m$, we have:
 \begin{itemize}
\item
$D^{'}\subseteq \bigcup\limits_{s=0}^{k-1}D_{1}(P_{{i}_{s}})$, and
\item the unit disk  graph with vertices $P_{i_{0}},P_{i_{1}},\dots
,P_{i_{k-1}}$ is connected.
\end{itemize}

\vskip0cm\noindent We note that event ${\cal K}_m$ implies that
the random unit disk graph which is formed from the vertices $P_1,
P_2,...,P_m$ has a $k$-point connected dominatng set. With this
notation we can state
\begin{theorem}
\label{kpts}
 There is a positive
constant $\alpha <1$ and a positive constant $m_{k}$ such that,
for all $m>m_k$, $\Pr({\cal K}_{m})> 1-4\alpha^{m}.$
\end{theorem}
\begin{proof}
Choose points $z_{0},z_1,z_2$ as in the proof of Lemma
\ref{ksectors}.
 If $z$ is any  point in $D_{\delta}(z_{s}), $ then for all
$y\in S_{s}$, $d(z,y)\leq d (z,z_{s})+d(z_{s},y)\leq \delta +\rho
<1.$
 Let
${\cal E}_{s}$ be the event that {\em none} of the $m$ random
points $P_1,P_2,\dots ,P_{m}$ lies in $D_{\delta}(z_{s}).$ Then
\begin{equation}
\Pr({\cal E}_{s})=\biggl(1-
 {{\rm Area}(D_{\delta}(z_{s})\bigcap
{\cal Q})\over {\rm Area}(D^{'})} \biggr)^{m}.
\end{equation}
Note that  ${\rm Area}(D_{\delta}(z_{s})\bigcap {\cal Q})\geq
{1\over 4}{\rm Area}(D_{\delta}(z_{s}))={\pi \delta^{2}\over 4},$
and that Area($D^{'})\leq $ Area($D_{1}(p))={\pi}.$ If we let
$\alpha=1-{\delta^{2}\over 4}=.998\dots$, then $\alpha <1,$ and
for $s=0,1,2,$
\begin{equation}
\label{bound1} \Pr({\cal E}_{s})\leq \alpha^{m}.
\end{equation}
It follows from(\ref{bound1}) that $\Pr({\cal K}_m)\geq
1-3\alpha^m$ since ${\cal E}_0^c\cap {\cal E}_1^c\cap {\cal
E}_2^c\subseteq {\cal K}_m$, and the proof is complete if $k=3$.

   Now suppose that
$k>3$,  and let $Y$ be the number of the $m$ random points that
lie in $D_{\rho}(z_{2})\bigcap {\cal Q}.$ Since $\{Y\geq
k\}\cap(\bigcap\limits_{s=0}^{2}{\cal E}_s^c)\subseteq {\cal
K}_m$, we have  \begin{equation} \label{complements} \Pr({\cal
K}_{m})\geq 1- \Pr(\bigcup\limits_{s=0}^{2}{\cal E}_{s})-
\Pr(Y<k).
\end{equation}
 But $Y$ has a binomial~$(m,\tilde{p})$ distribution, where
\begin{equation}
\tilde{p}={{\rm Area}(D_{\rho}(z_{s})\cap {\cal Q})\over {\rm
Area} (D^{'})}> {\pi\rho^{2}/4\over \pi}={3\over 16}.
\end{equation}
Hence, for all $m\geq k$,
\begin{equation}
\Pr(Y<k)=\sum\limits_{j=0}^{k-1}{m\choose j}\tilde{p}^{j}
(1-\tilde{p})^{m-j}
\end{equation}
\begin {equation}
 < m^{k}(1-\tilde{p})^{m-k}<({16m\over 13})^{k}\cdot
 ({13\over
16})^{m}.
\end{equation}
Since ${13\over 16} < \alpha, $ it follows  that,as
$m\rightarrow\infty$,
\begin{equation}
 \label{bound2}
 \Pr(Y<k)=o(\alpha^{m}).
\end{equation}
 Put(\ref{bound1}),(\ref{complements}), and (\ref{bound2})
 together to
conclude: there is a positive constant $m_{k}$ such that, for all
$m>m_{k},$
\begin{equation}
\Pr({\cal K}_{m}^{c})< 4\alpha^{m}.
\end{equation}

\end{proof}
\section{Analysis of Rule k}
In this section, we assume that $\ell_{n}=o(\sqrt{n})$ as
$n\rightarrow\infty$.
 Also, in this section, let $U_k=\sum_{i=1}^n I_i$ be a sum of
indicator variables where $I_i=1$ iff node $i$ is not included in
${\cal C}_k(V)$ under Rule k.
 Thus
Rule $k$ selects a dominating set ${\cal C}_{k}(V)$ having
$C_{k}(V)=n-{U_{k}}$ vertices, and it is desirable for ${U_{k}}$
to be large.
 Our goal in this section is to prove
that, for all $k>2,$ $E({U_{k}})\geq n-O(\ell_{n}^{2}).$
\par

Let $\lambda_{n}=n-\ell_{n}^{2},$ and let let $X_1, X_2, ..., X_n$
be independent, uniformly distributed random points in ${\cal Q}$,
namely the locations of vertices. (Here we are again simplifying
notation by writing $X_{i}$ instead of $X_{n,\ell_{n},i}$.) Let
$\rho_{i}$ be the number of neighbors of vertex $i$ having a
larger ID, i.e. the number of $j>i$ such that
$d({X}_{i},{X}_{j})\leq 1.$

\begin{lemma}
\label{bdegree} If $\ell_{n}=o(\sqrt{n}),$ then\
 $${\bf P}\Bigl( \rho_{i}<{(n-i)\pi\over 8\ell_{n}^{2}}\Bigr)\ \leq
 \exp( {-(n-i)\pi\over 32\ell_{n}^{2}}).$$
\end{lemma}

\begin{proof}
Let $|D_{1}({X}_{i})|={\rm Area}(D_{1}(X_{i})\bigcap {\cal Q})$ be
the area of the set of points in ${\cal Q}$ whose distance from
${X}_{i}$ is one or less. Thus $|D_{1}({X}_{i})|=\pi$ unless
${X}_{i}$ happens to fall near the border, and in all cases
$|D_{1}({X}_{i})|\geq {\pi\over 4}.$ Given $|D_{1}({X}_{i})|,$ the
variable $\rho_{i}$ has a
Binomial$\bigl(n-i,{|D_{1}({X}_{i})|\over \ell_{n}^{2}} \bigr)$
distribution. Therefore Chernoff's bound on the lower tail
distribution gives
$$
 {\bf P}\biggl(
\rho_{i}<{(n-i)\pi\over 8\ell_{n}^{2}}\biggr|\
|D_{1}({X}_{i})|\biggr)=$$

$$
{\bf P}\biggl( \rho_{i}<  {\pi\over 8|D_{1}({X}_{i})|}\cdot
{|D_{1}({X}_{i})|(n-i)\over \ell_{n}^{2}}\biggr|
|D_{1}({X}_{i})|\biggr)
$$

$$\leq \exp\biggl(
 -\Bigl(1-{\pi\over 8|D_{1}({X}_{i})|}\Bigr)^{2}\cdot
{|D_{1}({X}_{i})|(n-i)\over 2\ell_{n}^{2}}\biggr)
$$
$$\leq \exp\biggl(
{-(n-i)\pi \over 32 \ell_{n}^{2}}\biggr).
$$

\end{proof}

\begin{theorem} If $k>2$, then
$E_n(C_k)=O(\ell_{n}^{2}).$
\end{theorem}
\begin{proof}
Let ${\cal B}_{i}$ be the event that
 $\rho_{i}\geq {(n-i)\pi\over 8\ell_{n}^{2}}.$
By Lemma \ref{bdegree},
\begin{equation}
{\bf P}(I_{i}=1)\geq {\bf P}(I_{i}=1|{\cal B}_{i}){\bf P}({\cal
B}_{i})\geq {\bf P}(I_{i}=1|{\cal B}_{i})\Bigl(1-\exp\bigl(
{-(n-i)\pi \over 32 \ell_{n}^{2}})\Bigr)
\end{equation}
Now suppose that $i\leq  \lambda_{n}=n-\ell_{n}^{2},$ and observe
that
\begin{equation} \label{putsecond}
 {\bf P}(I_{i}=1|{\cal B}_{i})=
\sum\limits_{ v\geq {(n-i)\pi\over 8\ell_{n}^{2}}}{\bf
P}(I_{i}=1|\rho_{i}=v){\bf P}(\rho_{i}=v|{\cal B}_{i}).
\end{equation}
To estimate this, observe that
 \begin{equation}
 \label{putherefirst}
{\bf P}(I_{i}=1|\rho_{i}=v) =\int\limits_{{\cal Q}} {\bf
P}(I_{i}=1|\rho_{i}=v, {X}_{i}=\vec{x})f_{{X}_{i}}(\vec{x}|
\rho_{i}=v)d\vec{x}
 \end{equation}
where $f_{{X}_{i}}(\vec{x}| \rho_{i}=v)$ is the conditional
density of ${X}_{i}$ on the square ${\cal Q}$ {\sl given} that
$\rho_{i}=v.$ For $v> {(n-i)\pi/8\ell_{n}^{2}},$ Theorem
\ref{kpts} yields
\begin{equation}
{\bf P}(I_{i}=1|\rho_{i}=v, {X}_{i}={x})\geq 1-4\alpha^{v}\geq
1-4\alpha^{(n-i)\pi/8\ell_{n}^{2}}.
\end{equation}
Putting this back into (\ref{putherefirst}) and then
(\ref{putsecond}), we get
\begin{equation}
 {\bf P}(I_{i}=1|{\cal B}_{i})\geq 1-4\alpha^{(n-i)\pi/8\ell_{n}^{2}},
\end{equation}
and therefore
\begin{equation}
 {\bf P}(I_{i}=1)\geq {\bf P}(I_{i}=1|{\cal B}_{i}){\bf P}({\cal B}_{i})\geq
 (1-4\alpha^{(n-i)\pi/8\ell_{n}^{2}})
 \Bigl(1-\exp(-{(n-i)\pi\over 32\ell_{n}^{2} })\Bigr)
\end{equation}
\begin{equation}
\label{penult}
 \geq
   1-4\alpha^{(n-i)\pi/8\ell_{n}^{2}}
-\exp(-{(n-i)\pi\over 32\ell_{n}^{2} }).
\end{equation}
Recall that $\lambda_{n}=n-\ell_{n}^{2}$, and that the foregoing
 estimates were valid for all $i\leq \lambda_{n}$. Putting $j=n-i$, we
 get
\begin{equation}
E({U}_{k})\geq \sum\limits_{i=1}^{\lambda_{n}}{\bf
P}(I_{i}=1)=\sum\limits_{i=1}^{\lambda_{n}}
 \biggl(1-4\alpha^{(n-i)\pi/8\ell_{n}^{2}}
-\exp(-{(n-i)\pi\over 32\ell_{n}^{2} })\biggr)
\end{equation}

\begin{equation}
\geq\lambda_{n}-4\sum\limits_{j\geq \ell_{n}^{2}}
\bigl(\alpha^{\pi/8\ell_{n}^{2}}\bigr)^{j}- \sum\limits_{j\geq
\ell_{n}^{2}} (\exp(-\pi/32\ell_{n}^{2})\bigr)^{j}
\end{equation}

\begin{equation}
=n-O(\ell_{n}^{2}).
\end{equation}
\end{proof}
\section{Lower Bound}
If a vertex $v$ has higher ID than any of its neighbors, then it
cannot be eliminated under Rule $k$. This simple observation is
the basis for
\begin{theorem}\label{lowerbound} If $\ell_{n}=o(\sqrt{n}),$ then, for all sufficiently large $n$,
 the  expected size of the Rule $k$ dominating set
is more than $\ell_{n}^{2}/4$.
\end{theorem}

\begin{proof}
Let $L_{k}=\sum\limits_{i=1}^{n}I_{i},$ where $I_i=1$ iff node $i$
has a higher ID that all the nodes in $D_{1}(X_i)$. Note that
$I_{i}=1$ iff the nodes $X_{i+1},X_{i+2},\dots ,X_{n}$ all fall
{\em outside} the disk $D_{1}(X_{i}).$ Therefore
\begin{equation}
{\bf P}(I_{i}=1)=(1-{|D_{1}(X_{i})|\over \ell_{n}^{2}})^{n-i}\geq
(1-{\pi\over \ell_{n}^{2}})^{n-i}
\end{equation}

Therefore
\begin{equation}
E(L_{n})\geq \sum\limits_{i=1}^{n}(1-{\pi\over
\ell_{n}^{2}})^{n-i}={\ell_{n}^{2}\over \pi}(1-(1-{\pi\over
\ell_{n}^{2}})^{n})= {\ell_{n}^{2}\over \pi}(1-o(1)).
\end{equation}
\end{proof}

\section{Optimality}

For this section,  $ \ell_{n}\leq \sqrt{n\over a\log n},$ where
$a$ is a constant greater than 9.
 It is easy to verify that,
 with asymptotic probability one, there {\sl exists} a
 CDS, $C_{rand},$\ having $O(\ell_{n}^{2})$ vertices: simply partition
 the square ${\cal Q}$ into
$\lfloor 3\ell_{n}\rfloor ^{2}$ equal-sized squares,each with
sides of length $s_{n}={\ell_{n}\over \lfloor
3\ell_{n}\rfloor}={1\over 3}+O({1\over \ell_{n}})$, and then pick
 one node  from each of these small squares. More explicitly, for
$0\leq i,j< \lfloor 3{\ell_{n}}\rfloor,$ let $Q_{i,j}=
\bigl\lbrace (x,y): is_{n}\leq x< (i+1)s_{n}\ {\rm and}\
js_{n}\leq x< (j+1)s_{n}\bigr\rbrace .$ Let ${\cal B}$ be the
event that each of the $\lfloor 3\ell_{n}\rfloor^{2}$ small
squares contains one or more nodes. By Boole's inequality,
\begin{equation}
{\bf P}({\cal B}^{c}) \leq 9\ell_{n}^{2}
 {\bf P} (Q_{1,1}{\rm \ is\ empty})
 =9\ell_{n}^{2}(1-{1\over \lfloor 3\ell_{n}\rfloor^{2}}
 )^{n}
\end{equation}

\begin{equation}
=9\ell_{n}^{2}\exp\bigl( -{n\over 9\ell_{n}^{2}} (1+O({1\over
\ell_{n}^{2}})\bigr)
\end{equation}

\begin{equation}
\label{bcomp}
 < {n\over \log n}e^{-\log n}=O({1\over \log n}).
\end{equation}
 Now given the vertices $V=\{X_1, X_2, ..., X_n\}$, we construct
$C_{rand}\subseteq V$ as follows: For each $1\leq i,j\leq \lfloor
3\ell_n\rfloor$, if $Q_{i,j}$ contains at least one vertex, then
select one  vertex $V_{i,j}$ uniform randomly from among the
vetices in $Q_{i,j}$, and include $V_{i,j}$ in $C_{rand}.$ Thus
$C_{rand}$ is a (random) set of at most $\lfloor 3
\ell_{n}\rfloor^{2}$ nodes. It can contain fewer nodes (possibly
as few as one), but with asymptotic probability 1, $C_{rand}$
 contains exactly $\lfloor
3 \ell_{n}\rfloor^{2}$
 vetices and is a CDS.

\par
It is worth pointing out that this existence argument cannot be
used in a straight-forward way as the basis for a localized
algorithm because the nodes do not know their own locations in the
network. One of the main advantages of the Rule k algorithm is
that a vertex makes its decision based on very limited
information, namely its list of neighbors and their lists of
neighbors.
\par
Nevertheless, the existence argument is useful for us because it
leads to a lower bound the size that a CDS can have. The following
argument was influenced by \cite{MBHRJ}. The appendix of
\cite{DaiWu} is also pertinent, but we do not see how to turn the
discussion there into a mathematically rigorous proof.

Theorem 6 below is based on from the following observation:  If
$v$ is any point in ${\cal Q}$, then at most 81 nodes of
$C_{rand}$ are in $D_{1}(v)$. In particular, if $C_{opt}$ is a
minimum sized CDS, and $v$ is a node in $C_{opt}$, then $N(v)$
includes at most 81 nodes of $C_{rand}.$ But $C_{opt}$ is a
dominating set; therefore every node in $C_{rand}$ must be in
$N(v)$ for at least one $v\in C_{opt}.$ We therefore have a lower
bound of the size of $C_{opt}:$ \commentout{ The nodes of
$C_{rand}$ are not clumped together since there is at most one
node of $C_{rand}$ for each small square. If $v$ is any point in
${\cal Q}$, then there are less than 81 nodes of $C_{rand}$ that
are in $D_{1}(v)$. In particular, if $C_{opt}$ is a minimum sized
CDS, and $v$ is a node in $C_{opt}$, then $N(v)$ includes less
than 81 nodes of $C_{rand}.$
       But $C_{opt}$ is a dominating set; therefore every node in $C_{rand}$
       must be in $N(v)$ for at least one $v\in C_{opt}.$
 We therefore have a lower bound of the size of
$C_{opt}$:}
\begin{equation} \label{optbound}
 |C_{opt}|\geq {1\over
81}|C_{rand}|.
\end{equation}
 Combining (\ref{optbound}) with (\ref{bcomp}), we get
 \begin{theorem}
\label{loweroptbd} Suppose  $a>9$, and  $\ell_{n}\leq
\sqrt{{n\over a\log n}}$ for all $n$.  Then there is a constant
 $B>0$ such that, for all $n>1$,
  $${\bf P}_{n,\ell_{n}}\biggl( |C_{opt}| < {1\over 10}{\ell_{n}^{2}}\biggr)
< {B\over \log n}.$$
 \end{theorem}
\begin{corollary}
$E(|C_{opt}|) =\Theta(\ell_{n}^{2})$
\end{corollary}
\begin{proof}
From (\ref{optbound}), we have
$$E(|C_{opt}|)\geq {1\over 81}E(|C_{rand}|)$$

$$\geq {1\over 81}{\bf P}\Bigl(|C_{rand}|=\lfloor
3\ell_{n}^{2}\rfloor\Bigr) \cdot \lfloor 3\ell_{n}^{2}\rfloor$$

$$  ={1\over 81}(1-O({1\over \log n}))\lfloor 3\ell_{n}^{2}\rfloor
    =\Theta(\ell_{n}^{2}).$$
\end{proof}

\section{Discussion}
In this paper we have analyzed Rule $k$ only for $k>2$. For $k<3,$
the analysis is different and quite a bit more complicated. The
analysis for that case is treated in a subsequent paper. Also,
here we have only analyzed the application of Rule $k$ on the
entire vertex set of ${\cal G}_n$. Clearly Rule $k $ could also be
used in conjunction with other heuristics in order to construct a
\lq \lq small\rq\rq CDS. For example, Wu and Li have proposed the
\lq\lq Marking Process\rq\rq, an algorithm for selecting  an
initial CDS ${\cal M}.$ They recommended that the Marking Process
be followed by Rules $1$ and $2.$  Dai and Wu subsequently
proposed the more general Rule $k.$
 The various Rules $1,2,3,\dots $ can be applied one after the
 other up to some largest $k$. Dai Li and Wu
 mark the nodes in the CDS, and with each new rule application,
the set of marked nodes shrinks.
\par
In the case where $\ell_n\to\infty$ and $\ell_n<\sqrt{n/3\log n}$,
it can be shown (see Appendix 1) that asymptotically nothing is
gained by applying  the Marking Process before applying  Rule $k$.
It may be possible to obtain further reductions in the size of the
heuristic CDS by successive applications of Rules $1,2,..,k$ as
proposed by Dai Li and Wu. However, the rigorous analysis of the
Dai Li and Wu heuristic is complicated due to dependence  between
the variables at the various stages in the analysis of the
heuristic. For example, it seems much harder to estimate
 $E(|{\cal C}_{k+1}({\cal C}_{k}(V))|)$ than it is to estimate  $E(|{\cal C}_{k+1}({V})|)$ (say).
Our analysis  only considered $E(|{\cal C}_{k}({V})|)$ for any
fixed $k>2$ and we have shown that in this case the average size
of ${\cal C}_{k}({V})$ is of the same order as the size of the
optimal CDS. So, even a simple application of Rule $k$  to the
entire vertex set $V$ produces, on average,
 a \lq\lq good\rq\rq CDS.

 \vfill\eject\noindent
 {\bf Acknowledgement} We thank Li Sheng and Harish Sethu
 helpful comments.

\vfill\eject \noindent {\bf \large Appendix 1: The Marking
Process} \vskip.25cm\noindent
 Wu and Li \cite{WuLi} proposed the following method for selecting an
 initial CDS ${\cal M}:$
 \vskip.2cm\noindent
{Marking Process:} {\sl a node is included in ${\cal M}$ iff it
has two neighbors that are not adjacent (i.e. not directly
connected by and edge). }

\vskip.2cm\noindent Suppose we apply the Marking Process to the
random graph ${\cal G}_{n}.$ Let $M=|{\cal M}|$ be the number of
vertices marked by the marking process. In this appendix, let
$I_{i} =1$ iff the $i$th vertex gets marked, i.e. vertex $i$ has
two non-adjacent neighbors. Let $I_{i}=0$ otherwise. Thus $M
=\sum\limits_{i=1}^{n}I_{i}$ is
 the number of marked vertices. Our goal is to establish the
  following asymptotic  estimate for the expected
value of $M.$
\begin{theorem}
\label{avgM}
     $ E(M)=  n-O\bigl(n\exp(-{n\over \ell_{n}^{2}})\bigr).$
\end{theorem}

\begin{proof}
Since the  $I_{i}$'s are identically distributed, we have
\begin{equation} \label{identdistrib}
 E(M)=n{\bf P}(I_{1}=1). \end{equation}
 It therefore  suffices to prove that
${\bf P}(I_{1}=0)= O\bigl(\exp(-{n\over \ell_{n}^{2}})\bigr).$
%%%%%%%%%%%%%%%%%%%%%%%%%%%%%%%%%%%%%%%%
For any $i$, and any $r>0$,  let $D_{r}(i)$ be the disk of radius
$r$ centered at the vertex labelled $i.$ If
 vertex 1 happens to fall near the boundary of ${\cal Q},$
 then it may happen that part of
$D_{1}(1)$ is not entirely contained in ${\cal Q}.$ But in any
case we can partition $D_{1}(1)$ into four quarter disks and
select one of the four quarter disks $K$ in such a way  $K$ is
contained in ${\cal Q}.$ If $\varphi$ is the axis of symmetry of
$K$, let $B_1$ be the set of points in $K$ whose distance from
$\varphi$ is greater than ${1\over 2}.$  Note that $B_1$ consists
of two disjoint components $B_{1}^{+},B_{1}^{-},$ and that the
distance from $B_{1}^{+}$ to $B_{1}^{-}$ is 1. Hence vertex $1$
will be marked if both $B_{1}^{+}$ and $B_{1}^{-}$ contain at
least one of the other $n-1$ vertices. Define ${\cal B}_{1}$  to
be the event that both $B_{1}^{+}$ and $B_{1}^{-}$ contain at
least one of the other $n-1$ vertices. In this section only,
define $\alpha$ to be the area of $B_{1}^{+}.$
 The probability that  $B_{1}^{+}$ contains none of the other $n-1$ nodes is
$ ({ {\ell_{n}^{2}-\alpha}\over \ell_{n}^{2}})^{n-1}.$ The same is
true of $B_{1}^{-}.$ Hence
\begin{equation}
{\bf P}(I_{1}=0)\leq 2({ {\ell_{n}^{2}-\alpha}\over
\ell_{n}^{2}})^{n-1}= O(e^{-n/\ell_{n}^{2}}).
\end{equation}

\end{proof}

\begin{corollary} ${\bf P}(M\not= n)=O(ne^{-n/\ell_{n}^{2}}).$
\end{corollary}
\begin{proof}
By Boole's inequality,

${\bf P}(M\not= n)= {\bf P}(I_{i}=0$ for some $i) \leq n{\bf
P}(I_{1}=0)= O({ne^{-n/\ell_{n}^{2}}}).$
\end{proof}

\par
 Now fix $k\geq 2$, and let $C_{k}=|{\cal C}_{k}({V})|$
 be the number of vertices in the CDS which is
constructed when Rule k is applied to all nodes in the network.
Let $C_{k}^{'}=|{\cal C}_{k}({\cal M})|$ be the number of vertices
in the CDS which is constructed when Rule k is applied to ${\cal
M}=$ the nodes marked by the marking process. Provided
$\ell_{n}\to\infty$ and $\ell_{n}<\sqrt{n/3\log n}$, the two
quantities rarely differ, so we have the following corollary to
Theorem \ref{avgM}:
\begin{corollary}
\ {$E(C_{k}^{'})\geq E({ C}_{k})-O(n^{2}e^{-n/\ell_{n}^{2}}).$}
\end{corollary}

\begin{proof}

\begin{equation}
\label{ewlbnd}
 E(C_{k}^{'}) \geq E(C_{k}^{'}\bigr| M=n){\bf P}_{n}(M=n).
  \end{equation}
 If $M=n$, i.e. if ${\cal M}={V}$ and
 {\sl
  all} nodes in the network
 are marked, then ${\cal C}_{k}({V})={\cal C}_{k}({\cal M}). $ Therefore
 $E(
  C_{k}^{'}\bigr|
 {M}=n)
=E(C_{k} |{M}=n),$ and
\begin{equation}
\label{difference} E({C}_{k}^{'}\bigr| M=n){\bf P}( M=n)
=E(C_{k})-E( C_{k}|M\not=n){\bf P}(M\not=n).
\end{equation}
Combining (\ref{ewlbnd}) with (\ref{difference}), we get
$$E({C}_{k}^{'}) \geq
       E({ C}_{k})-E({C}_{k}|{ M}\not=n){\bf P}({ M}\not=n)
$$
$$
       \geq
      E({ C}_{k})-n{\bf P}({ M}\not=n)\geq E({ C}_{k})-O(n^{2}e^{-n/\ell_{n}^{2}}).
$$
\end{proof}

\end{document}